\documentclass[usenatbib]{mn2e}
\usepackage{color} 

\usepackage{graphics}
\usepackage{epsfig}
\usepackage{natbib}

\begin{document}

\title[Properties of AGN selected by mid-IR Colours]
{Properties of AGN selected 
by their mid-IR colours: evidence for a physically distinct mode of black hole
growth}

\author [G.Kauffmann] {Guinevere Kauffmann\thanks{E-mail: gamk@mpa-garching.mpg.de}\\
Max-Planck Institut f\"{u}r Astrophysik, 85741 Garching, Germany}

\maketitle

\begin{abstract} 
We study the narrow emission line properties and stellar populations
of a sample of 1385 AGN selected to have 
strong excess emission at mid-infrared wavelengths 
based on comparing  Wide-field Infrared Survey Explorer W1-W2 band
colours with optical stellar absorption line indicators. Our goal is 
to understand whether the  physical conditions in the
interstellar medium of these objects differ from
those of local AGN  selected by their 
optical emission line ratios. To enable this comparison, we construct a    
{\em control sample} of 50,000 optically-selected AGN with the same 
redshifts that do not have  strong mid-IR excess emission. 
The mid-IR
excess and control samples differ strongly in [OIII] line luminosity,
ionized gas excitation mechanism, ionization state and electron density. 
We show that the radio-detected, mid-IR excess AGN
constitute the most luminous and highly ionized  AGN in the local
Universe and they contribute primarily to the growth of black holes in the most
massive galaxies. At least half of this black hole growth is occurring in
galaxies with recent starbursts. The morphologies of
these systems indicate that the starbursts have probably been triggered by
galaxy-galaxy mergers and interactions. The most luminous AGN in our mid-IR excess sample
have properties that are similar to the Type II quasars studied at
higher redshifts. In contrast, the control sample
constitute a class of lower ionization, less luminous AGN in more
quiescent galaxies that
contribute primarily to the growth of low mass black holes. 
\end{abstract}


\section {Introduction}

Our understanding of the formation paths of supermassive black holes is
currently rather sketchy. The Soltan argument (Soltan 1982) applied to the
luminosity function of quasars measured at different redshifts shows that
the summed emission from  optically-selected AGN yields an estimate of the
mass contained in ``fossil quasars'' that is remarkably close to the total
estimated mass density in supermassive black holes at the present day (Yu
\& Tremaine 2002).  The discovery of extremely luminous quasars at
redshifts greater than 6 (Fan et al 2001), when the Universe was less than a
billion years old, then implies that the most massive of  these black holes,
with $M_{\rm{BH}} > 10^9 M_{\odot}$, must have  formed over very short timescales. Cosmological models
of black hole growth from seeds of $10^3-10^4 M_{\odot}$ show that extremely
massive, high-redshift black holes cannot form unless they accrete at a
supercritical rates, i.e. not limited by radiation pressure of the gas, for
at least part of their formation history (Volonteri \& Rees 2005,2006).

Very rapid growth of black holes likely occurred primarily at relatively
early epochs.  The total emissivity from quasars peaks at redshifts $\sim
2-3$ (see Manti et al 2017 for a recent compilation of data). A census of
the emission from low redshift AGN shows that most black holes
in the local Universe are accreting at low rates;  only a minority of the lowest
mass black holes in the local Universe with masses
of around $10^6 M_{\odot}$ are currently growing at rates that are
consistent with a formation timescale of less than a Hubble time (Heckman
\& Kauffmann 2004).  However, the existence of samples of hundreds of
thousands of low redshift AGN makes it possible to search for
sub-populations of objects that may be  undergoing much more efficient
black hole growth and where the fuelling processes may be occurring in a different mode to the  
bulk of the population.

In recent work, Kauffmann (2018; hereafter Paper I) introduced a new technique to identify AGN
via their [3.4]-[4.6] $\mu$m (WISE W1-W2 band) colours and radio emission.
A simple procedure was introduced  to pull out a ``mid-IR outlier''
population based on a combination of three stellar population diagnostics:
the 4000 \AA\  break strength, the specific star formation rate SFR/$M_∗$,
and the H$\delta_A$ Balmer absorption line index. Most of these mid-IR
outliers do not have centrally peaked emission and the mid-IR emission
likely originates from dust spread throughout the galaxy and not from a
central torus-like structure.  However, in Paper I we  showed that  by requiring
that the mid-IR outliers be detected in the Very Large Array (VLA) FIRST
Survey (Radio Images of the Sky at Twenty-Centimeters; Condon et al 1991), a
reasonably clean sample of  AGN with centrally-peaked emission that are
also identifiable as AGN using optical emission line diagnostics can be
recovered.

Interestingly, the fraction of systems with clearly disturbed morphologies
and signatures of recent or ongoing mergers was also very high in this
sample (Figure 12 in Paper I).  Although mergers and interactions
between galaxies have long been hypothesized to be a means of channelling
gas to the centers of galaxies and triggering accretion onto central
supermassive black holes (Sanders et al 2008), clear observational evidence of a
merger-triggered mode of accretion has been lacking so far. Carefully
controlled studies show that once the star formation rates of AGN and
non-AGN control samples are matched, there is no evidence that  AGN have a
higher incidence of close neighbours or asymmetric light profiles (Li et
al 2008; Reichard et al 2009, Cisternas et al 2011, Villforth et al 2014).  
The high incidence of clearly disturbed galaxies
in our  sample of radio-detected, mid-IR outliers therefore motivates
further study of these objects to understand whether or not they form a
class of AGN distinct from the population selected by optical emission line
ratios.

In this paper, we study the narrow emission line properties of a sample of
1385 radio-detected, mid-IR outliers in the main spectroscopic sample
of the Sloan Digital Sky Survey  in order to understand the physical
conditions in the interstellar medium of these objects in more detail and
to compare these with conditions in `ordinary' emission-line selected AGN.
We also study the stellar populations of the stars near the centers of the
host galaxies of these objects using Lick index measurements of key stellar
absorption features that probe stellar ages, metallicities and  $\alpha$ to
Fe element abundance ratios.  We interpret our findings using
photo-ionization and stellar population synthesis models. Finally, we
examine the contribution of radio-detected, mid-IR outliers to the total
black hole growth in the local Universe as a function of  host galaxy
stellar mass and  star formation history, as well as the  ionization state
of the narrow-line region (NLR) gas in these systems.

This paper is structured as follows. In section 2, we review our sample
selection criteria as well as the measurements of the emission and
absorption line quantities used in this work. In section 3, we present a
series of emission line diagnostic diagrams useful useful for understanding
the physical conditions in the ionized gas in our samples. In section 4, we
study the stellar populations of the host galaxies of our radio-detected
mid-IR outliers as well as our control sample of AGN, and in section 5, we
examine the contribution of the mid-IR outliers to the total growth of
black holes at the present day. In section 6, we discuss similarity
and differences of these results to those obtained in past studies of Type
II quasars, compact steep spectrum radio sources and radio galaxies in the
IRAS catalogue. Section 7 presents a summary of our findings and a
discussion of future perspectives.

\section {Construction of the Samples} As described in detail in Paper I, 
all the samples are  constructed from a magnitude-limited sample of
533,731 galaxies located in the main contiguous area of the final Data
Release (DR7; Abazajian et al (2009) of the main spectroscopic survey carried
out by the  Sloan Digital Sky Survey in the northern Galactic cap. The
galaxies are selected to have $r<17.6$, $−24<M_r<-16$ and spectroscopically
measured redshifts in the range $0.001<z<0.5$. Here $r$ is the $r$-band
Petrosian apparent magnitude, corrected for Galactic extinction, and $M_r$
is the $r$-band Petrosian absolute magnitude, corrected for evolution and
K-corrected to its value at z=0.1.

This sample is then cross-correlated  with the AllWISE Source Catalog,
which contains astrometry and photometry for 747,634,026 objects detected
on the deep AllWISE Atlas Intensity Images. 533,612 galaxies 
from the original sample are detected
in both the W1 and W2 bands, i.e. there is a  99.97\% detection rate. Paper I 
introduces a method for selecting so-called red outliers in the
plane of SFR/$M_∗$ versus D$_n$(4000) space for galaxies with $\log$
SFR/$M_∗ > −11$ and in H$\delta_A$ versus D$_n$(4000) space for galaxies
with lower specific star formation rates. Galaxies were binned in these two
planes and the outliers were selected as galaxies with W1-W2 colours lying
above the 95th percentile points in each bin. 

Examination of the central
colour gradients of 996 of the nearest of these outliers indicated that
radio luminosity was the property most predictive of redder W1-W2 colours
near the center of the galaxy, as would be expected if the hot dust
emission originated mainly from a central torus. Radio
luminosities were obtained by cross-correlating the full DR7 galaxy sample with the
FIRST Survey 
with a positional matching tolerance of 3 arcsec (see Paper
I for more details). We note that  
only $\sim 1\%$ of the mid-IR  outlier sample are detected at radio wavelengths,
making the radio selection a critical aspect of our technique compared to
previous work. As we will show, the combined mid-IR/radio selection yields
a sample of AGN with optical emission line luminosities that are much higher
than a ``control'' sample of AGN selected without regard to their radio
or mid-IR properties. This is unexpected, because pure radio selection yields
samples that are biased towards the most massive elliptical galaxies with little ongoing
star formation and optical emission line activity (Best et al 2005a,b).    

In this paper, we work with two samples: \begin{itemize} \item {\em
Radio-detected, mid-IR excess AGN} A sample of 1386 galaxies with stellar
masses in the range $10.0 < \log \rm{M_*} < 11.8$, redshifts less than 0.25 and
with radio luminosities greater than $10^{22.5}$ Watts Hz$^{-1}$,  with
H$\alpha$ luminosities $\log L(H\alpha)> 5.5$ L$_{\odot}$ and with both
H$\alpha$ and H$\beta$ detected with $S/N > 3$. \footnote {Note that in Paper I, we
showed that 80\% of all radio-detected, mid-IR outliers
were classified as optical AGN on the Baldwin, Phillipps \& Terlevich
(1981) (BPT) diagram and thus had secure detections of [OIII] and [NII] in
addition to H$\alpha$ and H$\beta$.} Unless indicated otherwise, all
emission line and stellar absorption line properties are taken from
the MPA-JHU release of spectrum measurements for DR7
galaxies. These are based on methods described in Brinchmann et al (2004)
and Tremonti et al (2004). \item {\em A control sample of optical
AGN} The control sample is selected so as to remove AGN that have significant 
excess emission at mid-IR or radio wavelengths. We pull out a 
sample of 50,084 optically-selected AGN that  have W1-W2
colours that do not place them  into the ``mid-IR excess category'' and that are
also not detected in the FIRST  catalogue of radio sources. They are further selected to have 
stellar masses, redshifts and H$\alpha$ luminosities in the same range as
the radio-detected, mid-IR excess AGN  sample. 

We  note that a galaxy is
defined to be an optically-selected AGN according to the demarcation in
[OIII]/H$\beta$ and [NII]/H$\alpha$ emission line ratios 
given in Kauffmann et al (2003b): \begin {equation}
\log\rm{([O III]/H\beta) > 0.61/(\log([N II]/H\alpha)-0.05)+1.3}
\end{equation} \end{itemize} This cut yields a larger optical AGN sample than
that of Kewley et al (2001) and allows us explore AGN over a larger
dynamic range in properties such as ionization parameter.  
In the following sections, we compare and
contrast the emission and stellar absorption line properties of these two
samples.
Note that
there are in total about 121,000 galaxies in this stellar mass
and redshift range in our sample that are classified as AGN 
without mid-IR excess. It is the
H$\alpha$  luminosity and H$\beta$ S/N cuts that are  responsible for reducing 
the number of control AGN  down to $\sim 54,000$ objects,
i.e. we are excluding the AGN with the weakest emission lines from
this comparison. Many weak emission line objects are classified
as LINERs in BPT line ratio diagnostic diagrams, but it has been shown
that most of the weak emission
is likely to  originate from ionization of gas by evolved stars such as  planetary nebulae     
(Capetti \& Baldi 2011; Singh et al 2013; Belfiore et al 2016).
The cut on galaxies with radio detections removes a further $\sim 4000$ objects.
Since this is a small fraction of the total sample, we do not attempt to
implement radio luminosity thresholds that scale with properties of the
galaxy, such as star formation rate.

\section {Physical conditions in the interstellar medium deduced from
emission lines}

\subsection {Excitation mechanism} In Figure 1, we begin with a set of
classic emission line diagnostic diagrams from BPT and from Veilleux \& Osterbrock
(1987), which have been constructed to probe  whether the excitation source is
an AGN or a starburst. The first of these diagnostic diagrams, $\log$
[OIII]/H$\beta$ versus $\log$ [NII]/H$\alpha$ was used to select our
optical AGN ``control'' sample shown as black points in the top left panel
of the figure.  The red points show the radio-detected, mid-IR excess AGN
and the demarcation between AGN and star-forming galaxies is shown as a
green curve on the  diagram.  Note that to make this figure, we have
selected a random sample of 1400 out of the 50,085 control galaxies, so
that the density of red and black points is the same in each panel.

\begin{figure}
\includegraphics[width=90mm]{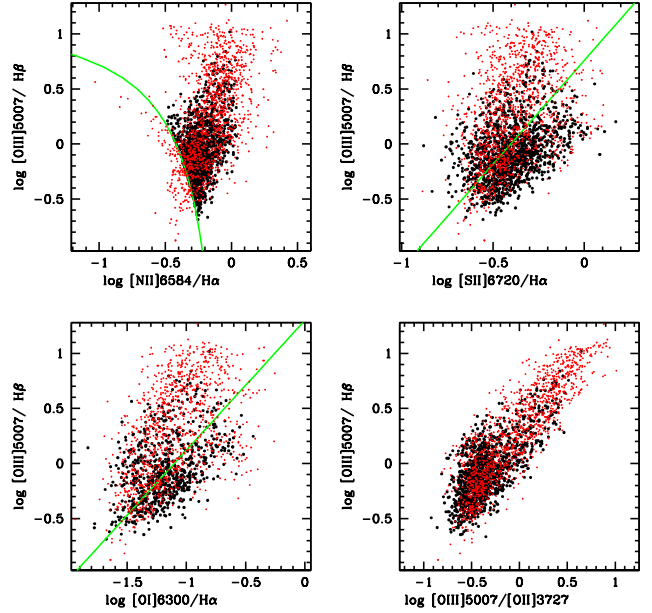}
\caption{ Badwin, Phillipps, Terlevich (BPT) diagrams showing the locations
of mid-IR excess AGN (red dots) and a random subset of control sample AGN 
(black dots) in the planes of [OIII]/H$\beta$ versus [NII]/H$\alpha$ (top left),
[OIII]/H$\beta$ versus [SII]/H$\alpha$ (top right), 
[OIII]/H$\beta$ versus [OI]/H$\alpha$ (bottom left), and
[OIII]/H$\beta$ versus [OIII]/[OII] (bottom right).
The green curve in the top left panel is the demarcation between 
AGN and star-forming galaxies from Kauffmann et al (2003b). The
green lines in the top right and botton left panels show the
demarcation between Seyferts and low-ionization AGN  given in Kewley et al (2006).
\label{models}}
\end{figure}

As can be seen, the majority of the black points cluster in the vicinity of
the green locus, indicating that a substantial fraction of the ionizing
photons for AGN in the control sample likely originate from HII regions in the galaxy.
The red points are spread much more widely over this diagram. In addition,  
there are many systems that have [OIII]/H$\beta$ radios close
to 10, which indicates that the ionizing photons in the mid-IR excess AGN
likely originate from  a source with a power-law spectrum at
UV/X-ray wavelengths, such as an accretion
disk.

The top-right and bottom-left panels show two additional diagnostic diagrams that were
extensively studied by Kewley et al (2006), who showed that there were two
clearly separated branches --  one corresponding to the
high-ionization Seyferts and another corresponding to lower ionization
sources, among them LINERs (low-ionization narrow emission line sources).
The boundaries between the two classes derived by Kewley et al  
are shown as green lines in the two
panels.  As can be seen, most galaxies in control sample cluster in the
low-ionization branch, and are located  
close to the region occupied by star-forming galaxies. The
radio-detected, mid-IR excess AGN lie mainly on the upper, high-ionization
branches in these two diagrams, and are more widely spread.  

In the bottom right panel, we plot mid-IR excess and control AGN is the plane of
$\log$ [OIII]/H$\beta$ as a function of  $\log$ [OIII]/[OII]. This diagram  probes
the ionization state of the narrow-line emitting gas in these systems. 
Here, the two samples superpose to form a single sequence, with the
radio-detected, mid-IR excess AGN offset towards higher ionization state
than the control sample.

\subsection {The spectral energy distribution of the radiation} The
intrinsic ionizing spectral energy distribution (SED) in AGN is uncertain.
In photo-ionization models of  AGN,  a power law radiation field $F_\nu
\propto \nu^{\alpha}$ is often assumed, with $\alpha$ a free parameter.
However, as discussed in the previous section, in many AGN, ionizing
photons may originate both from a combination of power-law sources and
young, massive stars.

The left panel of Figure 2 shows a slightly revised version of one of the
proposed SED diagnostic diagrams in a recent paper by Richardson et al
(2014), which combines a number of emission lines from the element oxygen
from different ionization states. Note that [OI], [OII] and [OIII] have
ionization energies of 13,6, 35.1 and 54.9 eV, respectively. [OI] is a
weaker line and is not detected in in all the galaxies in our two samples.
For this reason, the entire control sample, rather than just a random
subset, is plotted in the figure.  As seen in the previous figure, the
majority of galaxies in the control sample lie on a sequence bounded by low
[OIII]/[OII] ratios. Interestingly, the mid-IR excess sample scatters to
the higher [OIII]/[OII] ratios at {\em low} values of [OII]/[OI]
We note that 80\%  of the objects in the mid-IR excess
sample have high S/N [OI] detections, but only 50\% of the  control sample
galaxies are detected in [OI]. Most of the non-detected galaxies are low ionization
systems. This means that the actual shift of mid-IR excess objects towards higher
ionzation states is stronger than is already apparent from this diagram.

\begin{figure}
\includegraphics[width=90mm]{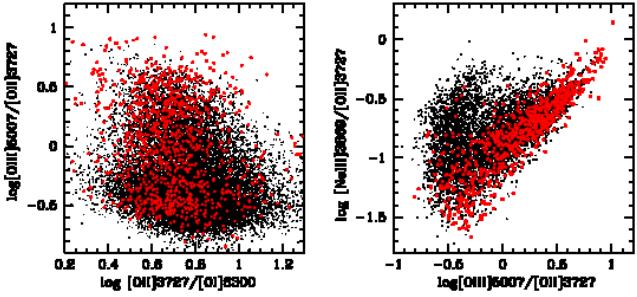}
\caption{ Two emission line rato diagrams that probe the spectral
energy distribution of the ionizing radiation in AGN. In the left panel,
mid-IR excess AGN (red) and control sample AGN (black) are shown in the
plane of [OIII]/[OII] versus [OII]/[OI]. In the right panel,
the two samples are shown in the plane of [NeIII]/[OII] versus
[OIII]/[OII].
\label{models}}
\end{figure}

Since the ionization energy  of [OI] is nearly identical to that of
hydrogen, in an ionization-bounded nebula [OI] is produced predominantly in
the ``partially ionized zone'',  wherein both neutral oxygen and free
electrons coexist (Ho 2008). In starburst galaxies, [OII] originates in
regions of the galaxy containing luminous young stars, while in powerful
Seyfert galaxies  [OIII] is often seen to map out cone-like structures of
outflowing, ionized gas (see for example Pogge 1988, Ferguson et al 1997).
One interpretation of the difference between the red and the
black points, therefore, may be the relative strength of the ionization
cones in the two classes of objects. We will come back to this hypothesis
later when we look at how the stellar populations of the underlying host
galaxy and the intrinsic luminosity of the AGN vary in different regions of
this diagram.

Levesque \& Richardson (2014) have proposed the [NeIII]/[OII] ratio as an
alternative ionization parameter diagnostic in star-forming galaxies. Neon
closely tracks the oxygen abundance in star-forming galaxies and HII
regions and is one of the principal coolants along with oxygen in the
ionized interstellar medium. [NeIII] has a higher ionization energy than
[OIII], so that the relation between [NeIII/[OII] and [OIII]/[OII] serves
as a diagnostic of changes in  the shape of the UV radiation field in
different galaxies. Finally, ionized neon is abundant over a broader range
of distances from the ionizing source in an ionized nebula than ionized
oxygen. This means that the [NeIII/[OII] ratio will be more sensitive to
very high ionization parameters than [OIII]/[OII].

The right panel of Figure 2 shows the location of the control AGN (black
points) and the radio-detected, mid-IR excess AGN in the plane of
[NeIII]/[OII] versus [OIII]/[OII]. We note that more than two thirds of the
mid-IR excess AGN have high S/N [NeIII] detections, but [NeIII/[OII]  can
only be measured for  only $\sim$ 10\% of the control sample. Nevertheless,
there are very clear differences in the locations of these two AGN
populations in this diagram. The control sample AGN appear to populate two
sequences: one where the [OIII]/[OII] ratio stays fixed around 0.5, but the
[NeIII]/[OII] ratio varies by an order of magnitude, and another where
[OIII]/[OII] and [NeIII]/[OII] correlate roughly linearly.  The mid-IR
excess AGN only populate the  second of these sequences.

In Figure 3, we plot the [NeIII]/[OII] ratio as a function of the AGN
luminosity as measured by the H$\alpha$ luminosity (left) and the [OIII]
luminosity of the galaxy. The mid-IR excess AGN are displaced towards
[OIII] luminosities that are, on average, an order-of-magnitude higher than those of 
the control sample AGN.  The displacement of the mid-IR excess
AGN to higher  [NeIII]/[OII] ratios is most
evident for mid-IR excess AGN with [OIII] luminosities greater than $10^8
L_{\odot}$. We will study  these very high-luminosity, high-ionization
parameter systems in more detail in section 6 of this paper.

\begin{figure}
\includegraphics[width=90mm]{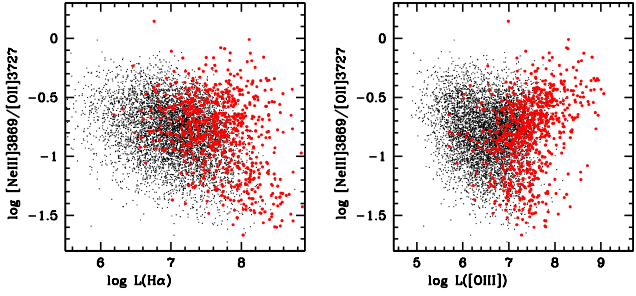}
\caption{ The [NeIII]/[OII] ratio is plotted as function of H$\alpha$
line luminosity (left panel) and [OIII]  line luminosity (right panel)
for mid-IR excess AGN (red) and control sample AGN (black). 
\label{models}}
\end{figure}

\subsection {Interstellar Medium Metal Abundances} Figure 4 shows ratios of
lines with similar ionization potentials, but from different elements,
leading to metal abundance sensitivity.  Both AGN samples span roughly the
same range in [NII]/H$\alpha$, [SII]/H$\alpha$ and [OI]/H$\alpha$, showing
that the metallicity of the ionized gas in these systems is about the same.

\begin{figure}
\includegraphics[width=90mm]{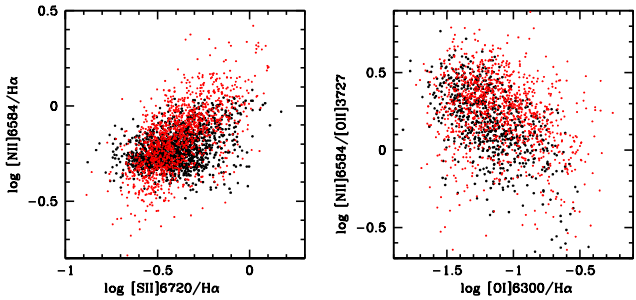}
\caption{ Two emission line rato diagrams that probe metal abundances
in the ionized gas.
In the left panel,
mid-IR excess AGN (red) and control sample AGN (black) are shown in the
plane of [NII]/H$\alpha$ versus [SII]/H$\alpha$. In the right panel,
the two samples are shown in the plane of [NII]/[OII] versus
[OI]/H$\alpha$.
\label{models}}
\end{figure}

\subsection {Interstellar Medium Electron Densities} Electron density,
$N_e$, is one of the key physical parameters characterizing an ionized
gaseous nebula.  The electron density in an ionized gaseous nebula can be
measured by observing the effects of collisional de-excitation on nebular
(forbidden) emission lines.  This is usually achieved by comparing the
observed intensities of lines emitted from two different energy levels of
nearly equal excitation energy from the same ion. If the two levels have
different radiative transition probabilities, then the relative populations
of the two levels will vary with electron density, as will the intensity
ratio of transitions emitted from them. In the optical wavelength region, a
commonly used density-diagnostic ratio  is
[SII]$\lambda$6716/$\lambda$6731, lower values of this ratio being
indicative of higher electron densities.

In Figure 5, we plot the [SII]$\lambda$6716/$\lambda$6731 doublet ratio as
a function of [OIII] line luminosity (left) and the [OIII]/[OII] ionization
parameter (right) for our two AGN classes. As seen previously, the mid-IR
AGN are clearly offset towards higher [OIII] luminosities and ionization
parameters compared to the control sample.  There is also a clear trend for
the electron densities to {\em increase} in more luminous AGN with higher
ionization parameters. Overall, $N_e$ is higher in the mid-IR excess sample
than in the control sample.

\begin{figure}
\includegraphics[width=90mm]{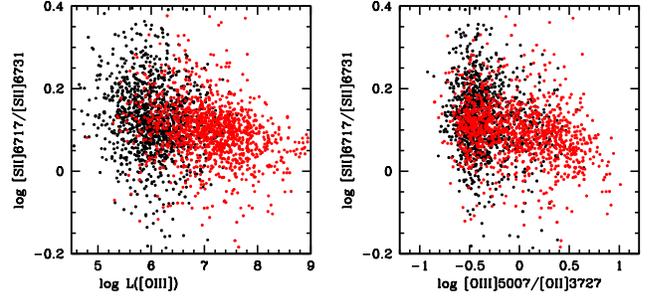}
\caption{ The electron density diagnostic [SII]$\lambda$6716/$\lambda$6731 doublet ratio
is plotted as a function of [OIII] line luminosity (left) and the [OIII]/[OII] ionization
parameter (right). Red points show mid-IR excess AGN and black points show control
sample AGN. 
\label{models}}
\end{figure}

\subsection {Dust in the Interstellar Medium} Given that the interstellar
medium electron densities in the mid-IR excess AGN are higher and that the
excess mid-IR radiation in these systems likely originates from dust grains
heated to high temperatures, one might ask whether radiation pressure from
dust grains plays a more important role in determining the emission line
properties of these systems compared to more ``normal'' AGN.

Dopita et al. (2002) developed models for the narrow line regions of AGN
where dust and the radiation pressure acting upon it provide the
controlling factor in moderating the density, excitation, and surface
brightness of photoionized NLR structures.  Additionally, photoelectric
heating by the dust is important in determining the temperature structure
of the models. In later work (Groves et al 2004b), the dusty models were
compared with simpler isochoric (constant density) dust-free models in a
series of line ratio diagram in an attempt to identify clear tests of the
models.

\begin{figure}
\includegraphics[width=90mm]{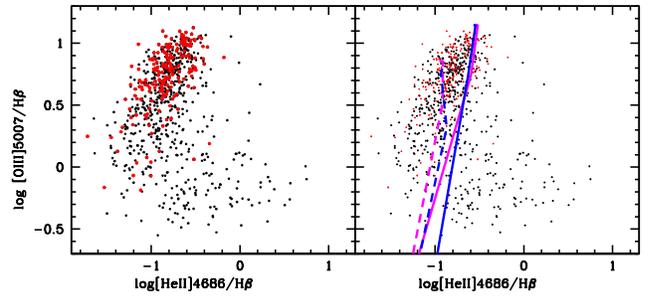}
\caption{An emission line diagnostic diagram that in sensitive to the effects of            
dust radiation pressure on the gas. {\em Left panel:} 
[HeII]/H$\beta$ is plotted as a function of   [OIII]/H$\beta$
for  mid-IR excess AGN (red points) and control
sample AGN (black points). {\em Right panel:} 
the dust-free isochoric models of Groves et al (2004a) are superposed
(2004) on the data. Magenta lines indicate the locus of models with
different ionization parameter for solar metallicity gas, while blue lines
are for 0.2 solar models.
Solid lines are
for ionizing radiation with  a flat power law spectrum ($\alpha=-1.2$, $F_{\nu} \propto
\nu^{\alpha}$), while dashed lines are for a steep power-law spectrum
($\alpha =-2$).  
\label{models}}
\end{figure}

In the left panel of Figure 6, we present a diagram that was first
introduced by Binette et al (1996) as a way of motivating the need for
photo-ionization models with complex gas density structure, as might be
expected in dusty regions of the galaxy. Groves et al (2004b)  showed that
the Dopita et al (2002) models where the gas density is controlled by
radiation-pressure from dust produce significantly higher [HeII]/H$\beta$
ratios when the ionization parameter is large compared to models without
dust. Note that because [HeII]$\lambda$4686 measurements are not available from
the MPA/JHU database, we have used measurements of this line provided by
the Portsmouth release of spectrum measurements (Thomas et al 2013).

The main result from the comparison of our two AGN samples is that the
mid-IR excess AGN and the control sample overlap almost exactly in
[HeII]/H$\beta$ at large values of [OIII]/H$\beta$.  In the right panel of
Figure 6, we superpose the dust-free isochoric models of Groves et al
(2004a) on the data. Magenta lines indicate the locus of models with
different ionization parameter for solar metallicity gas, while blue lines
are for 0.2 solar models.
The differences between the solid and dashed lines show the effect of
changing the shape of spectrum of the ionizing radiation -- solid lines are
for a flat power law spectrum ($\alpha=-1.2$, $F_{\nu} \propto
\nu^{\alpha}$), while dashed lines are for a steep power-law spectrum
($\alpha =-2$) with less contribution at shorter wavelengths.  As can be
seen, the extremely simple dust-free models with constant density are able
to fit the data remarkably well at large values of [OIII]/H$\beta$. We thus
conclude that there is no evidence that dusty NLR models provide an
explanation for differences in the emission line properties of our mid-IR
excess and control AGN samples.

In the data, the most extreme values of [HeII]/H$\beta$ are found for AGN
in the control sample and they occur at low values of [OIII]/H$\beta$. None
of the Groves et al photo-ionization models pass through this region of
parameter space. This may indicate that some other gas heating mechanism
may be at work in these objects.

\section {The
physical origin of the high ionization gas in mid-IR excess AGN}

In the previous section, we showed that the ionized gas in the mid-IR
excess AGN and in the control sample differ markedly in luminosity,
excitation mechanism, ionization state and electron density, but that the
ionized gas metal abundances are quite similar in the two samples and that
there is no evidence that radiation pressure from dust plays a more
important role in regulating the densities of the narrow-line regions of
the mid-IR excess objects.  In this section, we examine the stellar
populations and the radio properties of the two AGN classes in detail in
order to understand whether the presence or absence of  young, massive
stars and/or radio jets play any role in the observed differences.

\subsection {Clues from stellar absorption lines}
In this section we
examine relations between the following stellar absorption line diagnostics:
\begin{enumerate} \item The 4000 \AA\ break, D$_n$(4000). The break
occurring at 4000 \AA\ is the strongest discontinuity in the optical
spectrum and arises because of the accumulation of a large number of
spectral lines in a narrow wavelength region. In hot stars, the elements
are multiply ionized and the opacity decreases, so the 4000\AA\ break will
be small for young stellar populations and large for old, metal-rich
galaxies. We use the definition of the break defined in Balogh et al.
(1999) as the ratio of the average flux density F$_{\nu}$ in the bands
3850-3950 and 4000-4100 \AA\ and we will denote this index as D$_n$(4000).
\item Strong H$\delta$ absorption lines arise in galaxies that experienced
a burst of star formation that ended 0.1-1 Gyr ago. Worthey \& Ottaviani
(1997) defined an H$\delta_A$ index using a central bandpass bracketed by
two pseudo-continuum bandpasses. Kauffmann et al (2003a) and Kauffmann
(2014) showed that the location of a galaxy in the plane of H$\delta_A$
versus D$_n$(4000) could be used to distinguish galaxies that have
experienced a recent burst of star formation from galaxies that have had
continuous star formation histories over the past 1-2 Gyr.  \item The total
metallicity-sensitive index [MgFe]'.  Several studies have addressed the
dependence of Lick index strengths on changes in the relative ratios of
heavy elements (e.g. Tripicco \& Bell 1995;  Thomas, Maraston \& Bender
2003) These studies have led to the identification of composite $\rm{Mg+Fe}$
indices, which are sensitive to total metallicity (i.e.  the fraction by
mass of all elements heavier than helium over the total gas mass) but show
little sensitivity to $\alpha$/Fe (i.e. the ratio of the total mass of α
elements to the mass of iron). In this work, we use \begin{equation}
\rm{[MgFe]'= \sqrt{Mgb(0.72 Fe5270 + 0.28 Fe5335)}} \end{equation} as
proposed by Thomas et al. (2003).  \item The magnesium-sensitive index Mgb.
The location of a galaxy in the plane of Mgb versus [MgFe]' should then
provide an indication of the $\alpha$-enhancement of the stellar
population. \end{enumerate}

Figure 7 shows the location of our two classes of AGN in the planes of
$H\delta_A$ versus D$_n$(4000), Mgb versus D$_n$(4000), [MgFe]' versus
D$_n$(4000) and [MgFe]' versus Mgb. The mid-IR excess AGN are seen to
populate regions of these planes that are largely empty of AGN from the
control sample. The first panel shows that the mid-IR excess AGN include
many galaxies  with with low values of D$_n$(4000) and high values of
H$\delta_A$ that have likely experienced recent bursts of star formation.
We will show quantitatively in Section 5 the degree to which starburst
activity is boosted in this population compared to the control sample. The
second and third panels show that these bursty galaxies have low
metallicities. The fourth panel shows that the Mgb index is  somewhat
stronger at a fixed value of [MgFe]' for the mid-IR excess AGN, indicating
that their stellar populations are more enhanced with $\alpha$-elements.
Again, this is exactly what would be expected if some of these systems have
experienced recent starbursts.

\begin{figure}
\includegraphics[width=90mm]{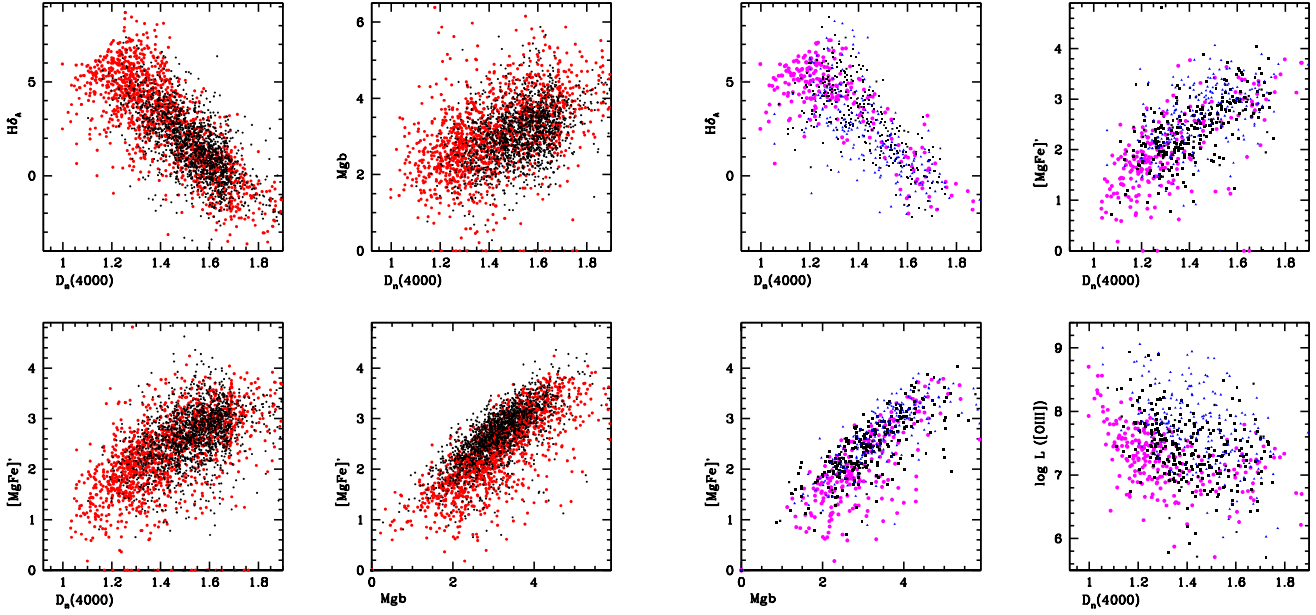}
\caption{Stellar aborption line index diagrams comparing the stellar population
properties of the host galaxies of mid-IR excess AGN (red points) with control
sample AGN (black points). Galaxies are shown in the planes of H$\delta_A$
versus D$_n$(4000) (top left), Mgb versus D$_n$(4000) (top right),
[MgFe]' versus D$_n$(4000) (bottom left) and [MgFe]' versus Mgb (bottom right). 
\label{models}}
\end{figure}

\subsection {Relationship between star formation history and ionization
state} We now examine how the ionization state of the gas in the mid-IR
excess sample varies in the 4 stellar population diagrams shown in Figure
7. In Figure 8,  we split the sample into 3 different ranges in $\log \rm{[NeIII]/[OII]}$: AGN
with $\log \rm{[NeIII]/[OII]>-0.6}$ are colour-coded blue, those with $-0.9<
\log \rm{[NeIII]/[OII]}<-0.6$  are colour-coded in black  and those with $\log
\rm{[NeIII]/[OII]}<-0.9$ are colour-coded in magenta. We see from this figure 
that the galaxies with the lowest [NeIII]/[OII] values coloured in
magenta have  young, metal-poor stellar populations. 

\begin{figure}
\includegraphics[width=90mm]{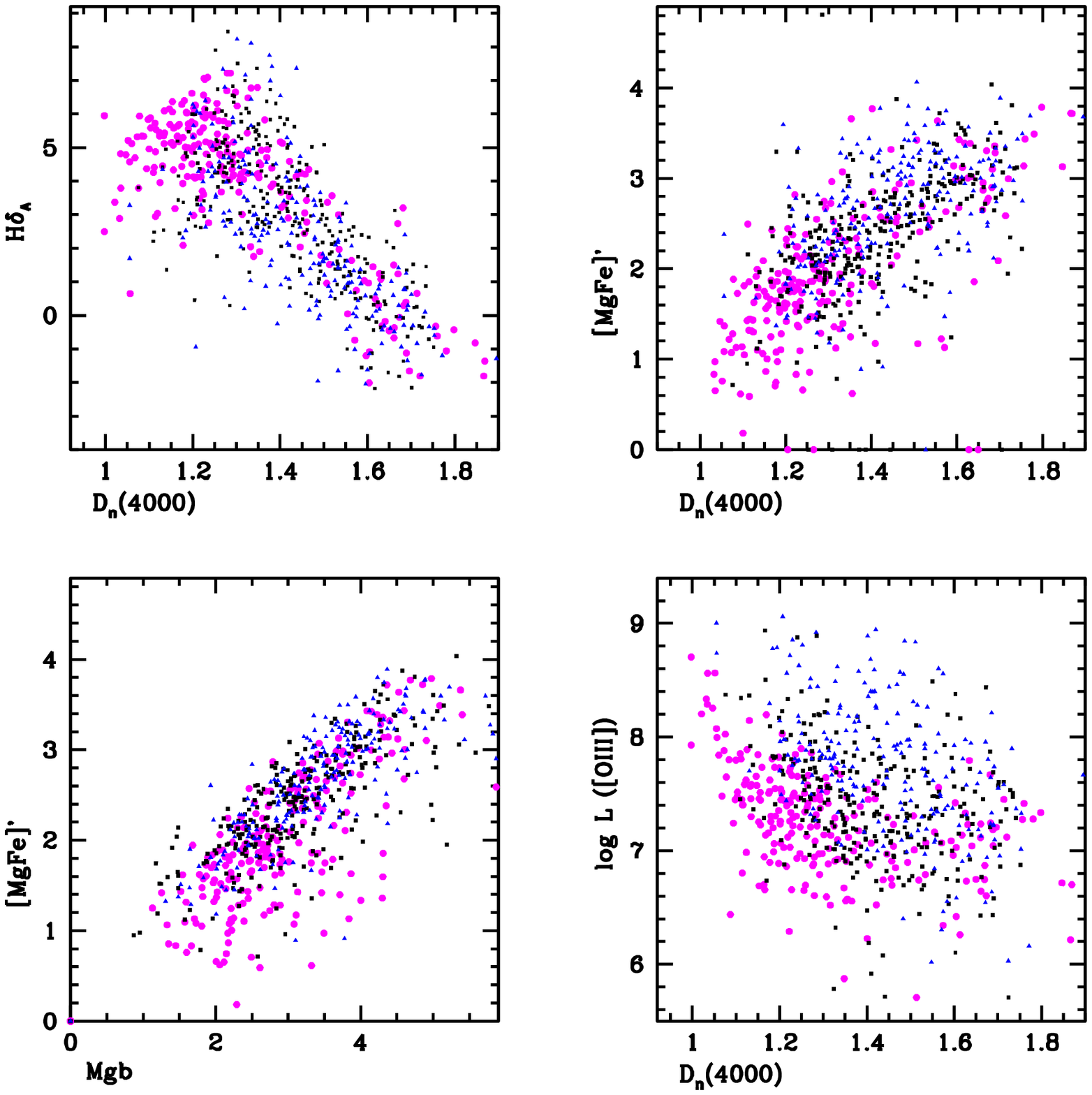}
\caption{The mid-IR excess AGN plotted in Figure 7 are colour-coded
according to [NeIII]/[OII]. AGN  with $\log \rm{[NeIII]/[OII]>-0.6}$ are plotted as blue
traingles, those with $-0.9<
\log \rm{[NeIII]/[OII]}<-0.6$ as black squares  and those with $\log
\rm{[NeIII]/[OII]}<-0.9$ as  magenta circles. 
\label{models}}
\end{figure}

In Figure 9, we expand upon this result by colour-coding AGN in the plane
of [NeIII]/[OII] versus [OIII]/[OII] according to their D$_n$(4000) values.
Magenta points indicate galaxies with 1.6$<$D$_n$(4000)$<$1.8 , red points
indicate galaxies with 1.4$<$D$_n$(4000)$<$1.6, green points galaxies with
1.2$<$D$_n$(4000)$<$1.4 and blue points galaxies with 1.0$<$D$_n$(4000)$<$1.2.
Results are shown separately for control sample AGN and for 
mid-IR excess AGN. We note that galaxies that are colour-coded blue
have D$_n$(4000) values that cannot be explained unless
their present-day star formation rates are  elevated  compared to their
past average ones, i.e. they are currently experiencing
a burst of star formation (Kauffmann 2014). 

Very similar patterns of
variation in D$_n$(4000) are seen for the control sample AGN and for the
mid-IR excess AGN in Figure 9. Galaxies with  [NeIII]/[OII] values greater than $\sim
0.1$ have predominantly old stellar populations. Starburst galaxies (blue
points) have low  [NeIII]/[OII] values in both samples. The relation
between D$_n$(4000) and [OIII]/[OII] is more complex. Galaxies with old
stellar populations  occupy a wide range in [OIII]/[OII] ratio in both
samples. It is only the young starburst galaxies that can clearly be
distinguished by their low [OIII]/[OII] values.

\begin{figure}
\includegraphics[width=90mm]{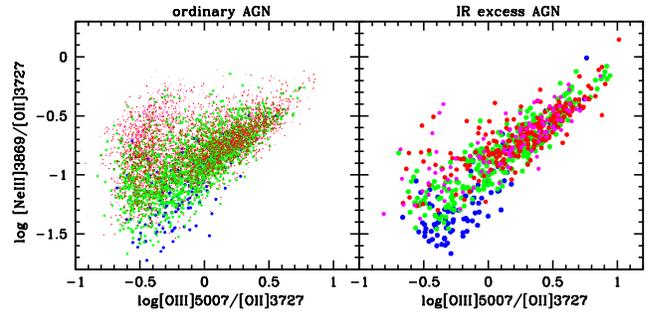}
\caption{AGN in the plane
of [NeIII]/[OII] versus [OIII]/[OII] are
colour-coded according to their D$_n$(4000) values.
Magenta points indicate galaxies with 1.6$<$D$_n$(4000)$<$1.8 , red points
indicate galaxies with 1.4$<$D$_n$(4000)$<$1.6, green points galaxies with
1.2$<$D$_n$(4000)$<$1.4 and blue points galaxies with 1.0$<$D$_n$(4000)$<$1.2.
Results are shown separately for control sample AGN (left) and for
mid-IR excess AGN (right).
\label{models}}
\end{figure}

\subsection {Relationship between AGN luminosity and ionization state} In
Figure 10, we colour-code AGN in the plane of [NeIII]/[OII] versus
[OIII]/[OII] according to their [OIII] line luminosities.  Black points
indicate AGN with $5<\log\rm{L[OIII]}<6$ , red points indicate AGN with $6<\log
\rm{L[OIII]}<7$, green points AGN with $7<\log\rm{L[OIII]}<8$ and blue points AGN
with $8<\log\rm{L[OIII]}$. Once again, the control sample is shown in the left
panel and the mid-IR excess AGN in the right panel. Because the cuts on AGN
luminosity have been  made using the [OIII] line, it is natural to expect that
high luminosity and low luminosity AGN will segregate along the x-axis of
the plot. Interestingly, there is only a weak relation between
[NeIII]/[OII] ratio and AGN luminosity  for the control sample, but in the
mid-IR excess sample we clearly see that the highest luminosity systems
with $\log\rm{L[OIII]}>8$ have very high [NeIII]/[OII]. Moreover, these very
high luminosity systems lie on a remarkably tight locus in the
[NeIII]/[OII] versus [OIII]/[OII] plane.

\begin{figure}
\includegraphics[width=90mm]{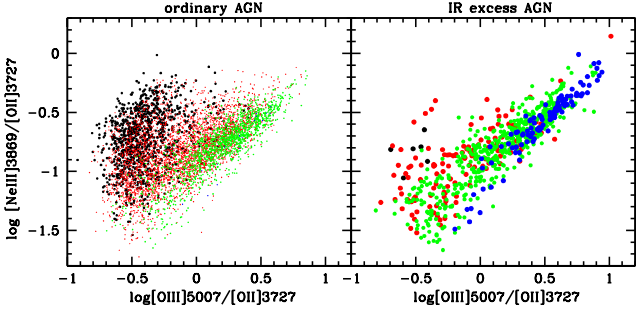}
\caption{AGN in the plane
of [NeIII]/[OII] versus [OIII]/[OII] are
colour-coded according to [OIII] line luminosity.  
 Black points
indicate AGN with $5<\log\rm{L[OIII]}<6$ , red points indicate AGN with $6<\log
\rm{L[OIII]}<7$, green points AGN with $7<\log\rm{L[OIII]}<8$ and blue points AGN
with $8<\log\rm{L[OIII]}$. Results are shown separately for control sample AGN (left) and for
mid-IR excess AGN (right). 
\label{models}}
\end{figure}

In the previous subsection, we found that AGN with very high [NeIII]/[OII]
have predominantly old stellar populations in their central regions as
measured by their D$_n$(4000) break strengths. Note that at the mean
redshift of the sample ($z \sim 0.1$), the 3 arcsec SDSS fibre spectrum
subtends a physical region of radius 2.8 kpc at the center of the galaxy,
i.e. the spectrum is indicative of the mean age of the stellar population
of {\em all the stars in the bulge}.  It is still possible that a smaller
population of young, very massive stars concentrated in the vicinity of the
central supermassive black hole is responsible for the ionization of
[NeIII], as has been found in the bulge of our own Milky Way     
(Serabyn, Shupe \& Figer 1998). 

To explore this hypothesis in more detail, we turn to the stellar
photo-ionization grids of Levesque \& Richardson (2014), which predict the
relation between [NeIII]/[OII] and [OIII]/[OII] for the ionized gas in the
vicinity of a population of young massive stars over a very large range in
ionization parameter. The models in Levesque \& Richardson (2014) are an
update of the Levesque  et al (2010) models, which span a range of
ionization parameters $10^7$cm s$^{-1}$ $< q< 4 \times 10^8$ cm s$^{-1}$
chosen to agree with observed ionization parameters in local starburst
galaxies. Richardson et al (2013) extended these models from $q$ values of
$6 \times 10^8$ cm s$^{-1}$ up to the theoretical maximum of $q_{max}=c$.
This was done in order to provide models that could better fit the emission
line properties of strongly star-forming galaxies at high redshifts.

In the left panel of  Figure 11, we plot the relations between
[NeIII]/[OII] and [OIII]/[OII] given in Table 1 of Levesque \& Richardson
(2014) together with the data for the most luminous mid-IR excess AGN with
$\log\rm{L[OIII]}>8$. Model results are shown for 5 different metallicities
(Z=0.001, cyan; Z=0.004, green, Z=0.008, black, Z=0.020, red; Z=0.040,
magenta). As can be seen, models with  super-solar metallicities are
required to explain the data. We note Galactic center M-giants have
super-solar metallicities (Ryde \& Shultheis 2015), so this conclusion is
not unreasonable. The right panel of Figure 11 shows [NeIII]/[OII] as a
function of $q$ for the different models. Unfortunately, Levesque \&
Richardson did not extend their super-solar model to very high values of
$q$. Nevertheless, we can deduce from the plot that $q$ values  in the
range $\sim 10^9-3\times 10^9$cm s$^{-1}$ would likely be needed to match
the AGN with the highest [NeIII]/[OII] values in our sample. Whether this
is achievable with normal massive, young stars at the centers of these
galaxies remains an unanswered question. The alternative explanation is
that harder spectrum radiation from a central accretion disk ionizes the
gas, but a different explanation for the observed tight correlation between
[NeIII]/[OII] and [OIII]/[OII] would then have to be found.

\begin{figure}
\includegraphics[width=90mm]{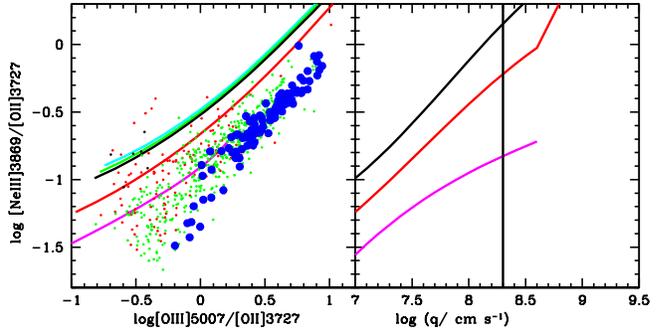}
\caption{ {\em Left panel:} The relations between
[NeIII]/[OII] and [OIII]/[OII] given in Table 1 of Levesque \& Richardson
(2014) are plotted together with the data. The most luminous mid-IR excess AGN with
$\log\rm{L[OIII]}>8$ are highlighted as larger filed blue circles. 
Model results are shown for 5 different metallicities
(Z=0.001, cyan; Z=0.004, green, Z=0.008, black, Z=0.020, red; Z=0.040,
magenta). {\em Right panel:} [NeIII]/[OII] is plotted as a
function of the ionization parameter  $q$ for the different models.
\label{models}}
\end{figure}

\subsection {Relationship between radio luminosity and ionization state} In
Figure 12, we plot the [OIII] line luminosities and the [NeIII]/[OII]
ratios as a function of the radio luminosities of our mid-IR excess AGN.
We find no correlation. There is thus no evidence that a
synchrotron-emitting radio jet is responsible for energizing the gas in
these systems.

\begin{figure}
\includegraphics[width=90mm]{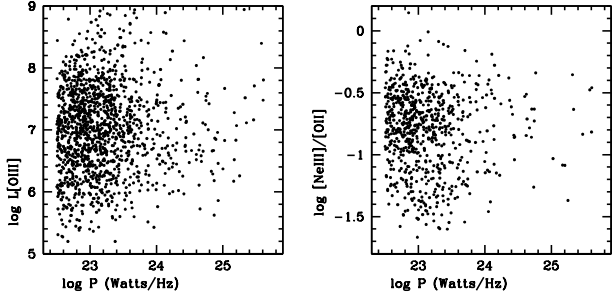}
\caption{ [OIII] line luminosities (left)  and  [NeIII]/[OII]
ratios (right) are plotted as a function of the radio luminosities of
the mid-IR excess AGN. 
\label{models}}
\end{figure}

\section 
{Contribution of different modes of black hole growth} We began this
analysis with the hypothesis that the mid-IR excess AGN and the control
sample AGN constituted two disjoint modes of black hole growth in the local
Universe. This hypothesis was motivated by results in Paper I 
showing that the fraction of merging/interacting galaxies among the mid-IR
excess population was surprisingly large.

In this paper, further strong evidence that the two populations are
disjoint emerges from the fact that the typical [OIII] luminosities are
very different for the two classes of AGN  We now illustrate this point
quantitatively. Because our two AGN samples are drawn from a complete
$r$-band magnitude limited sample of galaxies, it is simple to calculate
the relative number densities of AGN in the two classes by weighting each
galaxy by $1/V_{max}$ where $V_{max}$ is the total volume over which the
galaxy can be detected in the survey.  The left hand panel of Figure 13
shows the logarithm of the ratio of the number density of mid-IR excess AGN
to that of AGN in the control sample as a function of [OIII] line
luminosity. The black points show results for the measured [OIII] line
fluxes without any extinction correction. The red points show the effect of
correcting the [OIII] line flux for extinction using the measured Balmer
decrement H$\alpha$/H$\beta$. \footnote{To correct for extinction, we
adopt the A$_V$ = 1.9655R$_V$ log(H$\alpha$/H$\beta$/2.87), where we assume
$R_V = 3.1$ and adopt the extinction curve given in equation (3) of
Wild et al (2007) with $\mu=0.3$.} As can be
seen, for low luminosity AGN with [OIII] line luminosities $\sim 10^6 -
10^7 L_{\odot}$, control sample AGN are 30-1000 time more numerous than
mid-IR excess AGN. For [OIII] line luminosities greater than $10^8
L_{\odot}$, the number density of mid-IR excess AGN begins to exceed the
number density of control sample AGN.

The right-hand panel of Figure 13 shows the ratio of the integrated [OIII]
luminosity in the two classes as a function of the stellar mass of the host
galaxy. This plot follows the approach taken in Heckman \& Kauffmann
(2004), where the [OIII]$\lambda$5007 line was used as a proxy for black
hole accretion rate and the integral over the [OIII] luminosities of a
sample of galaxies represented the total contribution of the population to
black hole growth in the local Universe.  For galaxies with stellar mass
comparable to that of the Milky Way ($\log M_* =10.6$), we find that the
total [OIII] luminosity contributed by the mid-IR excess population is
around a quarter of that contributed by the control sample AGN. The
relative contribution from the mid-IR excess AGN
 increases with stellar mass and for galaxies that are 10
times more massive than the Milky  Way, mid-IR excess AGN contribute as
much integrated [OIII] luminosity as the control sample.

\begin{figure}
\includegraphics[width=90mm]{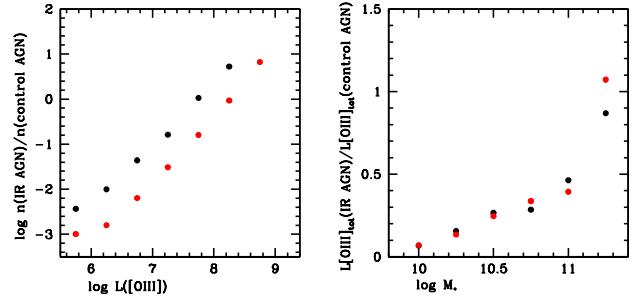}
\caption{ {\em Left panel:} The logarithm of the ratio of the number density of mid-IR excess AGN
to that of AGN in the control sample is plotted  as a function of [OIII] line
luminosity. 
{\em Right panel:} The ratio of the integrated [OIII]
luminosity in mid-IR excess AGN and control sample AGN  as a function of the stellar mass of the host
galaxy.    
In both panels, the black points show results for the measured [OIII] line
fluxes without any extinction correction. The red points show the effect of
correcting the [OIII] line flux for extinction using the measured Balmer
decrement. 
\label{models}}
\end{figure}

In Figure 14, we illustrate two more ways in which the mid-IR AGN and the
control AGN differ. In the top panels, the solid lines show how  the total
integrated [OIII] luminosity from AGN is partitioned as a function of
stellar mass of the host galaxy. Results for the control sample are shown
in the top left panel and for the mid-IR excess sample in the top right
panel. In both samples, most of the integrated [OIII] emission is located
in galaxies with masses close to that of the Milky Way (i.e. $\log M_* \sim
10.6$). The much larger number densities of control sample AGN at these
stellar masses shown in Figure 13 tells us that the {\em duty cycle} of
this mode of accretion is much longer.  This point can also be made by
looking at the contribution of galaxies that have undergone a recent burst
of star formation to the integrated [OIII] luminosity in both samples. This
is shown by the dashed curves in each of the top two panels. 

In order to
divide galaxies into bursting and non-bursting classes, we use a  grid of
model star formation histories fit to the measured values of D$_n$(4000)
and H$\delta_A$. The reader is referred to Kauffmann (2014) for more
details about this methodology. In summary the model library includes
models with continuous star formation histories,
models with ongoing bursts and models with  bursts that have occurred
between 0.2 and 2 Gyr in the past.  
Continuous models occupy a narrow locus between the regions of
the H$\delta_A$ versus D$_n$(4000)  diagram spanned by the two classes 
of burst models. Model galaxies with ongoing bursts
are displaced to lower H$\delta_A$ values at fixed 
D$_n$(4000), whereas model galaxies with a past burst
are displaced to higher H$\delta_A$.  We first search the continuous star formation library for the model that minimizes
$\chi^2$. We use the minimum $\chi^2$ value to  assess  
whether the continuous star formation
history probability has greater than 50\% probability of being  correct.
If not, we then search both the ongoing and past burst libraries for a new minimum
$\chi^2$ model. If the new minimum is smaller than that obtained for the
continuous libary, the classification as a starburst system is considered as
``secure''.  

A comparison of the dashed curves in the
top left and right panels of Figure 14, shows that the contribution of galaxies with
starbursts to the total [OIII] luminosity is a factor of 5-7 times higher
in the mid-IR excess sample than in the control sample.  Starburst galaxies
contribute approximately half the total integrated [OIII] luminosity in mid-IR excess
AGN compared to less than a tenth in the control sample.

Finally, the two bottom panels of Figure 14  show how  the total integrated
[OIII] luminosity from AGN is partitioned as a function of the [OIII]/[OII]
ratio for the two AGN classes.  The peak of the distribution function is 
shifted to [OIII]/[OII]  ratios that are a factor of 10 higher in the mid-IR
excess AGN compared to the control sample.  Once again, these results
illustrate the very strong dichotomy in the physical conditions in the gas
in these two classes of active galaxies.

\begin{figure}
\includegraphics[width=90mm]{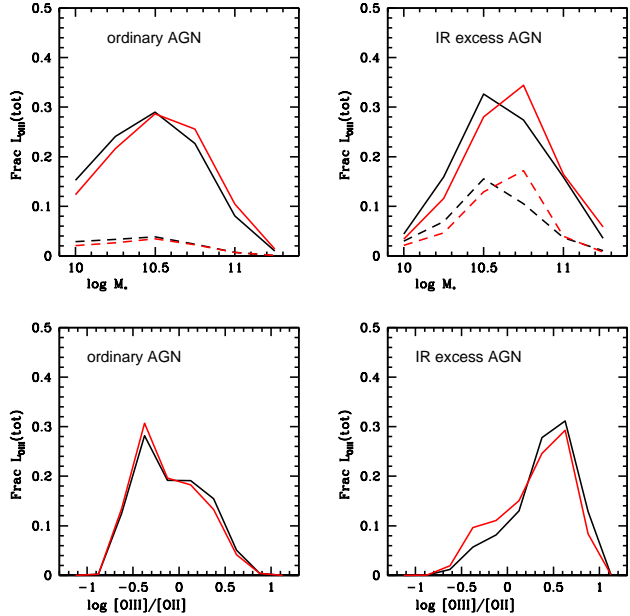}
\caption{ {\em Upper panels:} 
The solid lines show how  the total
integrated [OIII] luminosity from AGN is partitioned as a function of
stellar mass of the host galaxy. 
The dashed curves show the contribution of galaxies that have undergone a recent burst
of star formation. {\em Lower panels:} Partition of  the total integrated
[OIII] luminosity from AGN  as a function of the [OIII]/[OII]
ionization parameter.  Results for the control sample are shown   
in the left panels and for the mid-IR excess sample  in the right panels.
The black curves  show results for the measured [OIII] line
fluxes without any extinction correction. The red curves  show the effect of
correcting the [OIII] line flux for extinction using the measured Balmer
decrement. 
\label{models}}
\end{figure}

In summary, we conclude that our radio-detected, mid-IR excess AGN
constitute the most luminous and strongly ionized  AGN in the local
Universe and contribute primarily to the growth of black holes in the most
massive galaxies. At least half of this black hole growth is occurring in
galaxies with recent starbursts that have likely been triggered by
galaxy-galaxy mergers and interactions. In contrast, the control sample AGN
constitute a class of lower ionization, less luminous systems that
contribute primarily to the growth of black holes in low mass galaxies over
much longer timescales.

\section {Relation of mid-IR excess AGN sample to other samples from the
literature} In this section, we briefly discuss similarity and differences
of the results presented in this paper to those obtained in past studies of
a variety of low-redshift luminous AGN, including compact steep spectrum
radio sources,  radio galaxies with detections in the Infrared Astronomical
Satellite (IRAS) catalogue and Type II quasars. In addition, we clarify how our
sample differs from AGN samples selected only by their mid-IR colours.

\subsection {Compact steep-spectrum radio sources} The gigahertz
peaked-spectrum (GPS) and compact steep-spectrum (CSS) radio sources make
up significant fractions of the bright (centimeter-wavelength-selected)
radio source population (~ 10\% and ~ 30\%, respectively).  The GPS sources
are powerful ($\log \rm{P (1.4 Ghz)} > 25$ W Hz$^{-1}$) and  compact ($<$1 kpc).
The CSS sources are just as powerful, but are larger (1-20 kpc in size).
The GPS and CSS sources are believed to be the be the younger stages of
powerful large-scale radio sources.

The host galaxy morphologies of CSS sources have been studied by Gelderman
\& Whittle (1994) and evidence has been found for  diffuse linear features, such
as tidal tails, bridges, and shells, indicative of a recent interaction.
Gelderman \& Whittle present low-dispersion spectra (and
high-dispersion spectra of the region around [O III]$\lambda$5007) of a
sample of 20 CSS sources (both galaxies and quasars). The main result is  
that the CSS sources have relatively strong, high equivalent width,
high-excitation line emission, with broad, structured [O III] profiles.
They suggest that these properties are consistent with strong interactions
between the radio source and the ambient line-emitting gas.  Holt,
Tadhunter \& Morganti (2009) also find a preponderance of complex,
multi-component emission lines in compact radio galaxies, which they
interpret in terms of jet-cloud interactions.  Clear correlations are
observed between the total radio luminosities of the sources and their [OIII]
line luminosities, lending further support to this picture.

Our sample of radio-detected mid-IR excess sources is quite different.
Firstly, the radio luminosities of our objects are  all less than $10^{25}$
W Hz$^{-1}$ and as shown in Figure 12, there is no correlation between
radio and [OIII] line luminosity. Second, the emission lines in our objects
are narrow and there is no evidence for velocity shifts indicative of large
gas motions in the majority of our objects (see Paper I).

\subsection {Radio-excess IRAS galaxies} The selection of radio-excess IRAS
galaxies was described in a paper by Drake et al (2003). Objects were found
by cross-correlating the Parkes-MIT-NRAO 5 GHz radio source catalogue
with the IRAS Faint Source Catalogue. 
Objects having more than 5 times as much radio emission as
expected from the FIR-radio correlation followed by normal star-forming
galaxies were selected to form the radio excess sample.  The radio
luminosities are in the range $23.5<\log\rm{P(5 Ghz)}<26$, i.e.  more
luminous on average than our mid-IR excess sources, but less luminous than
the CSS and GPS sources discussed above.

Follow-up optical spectroscopy is discussed in a paper by Buchanan et al
(2006).  These authors find that the radio excess is an excellent indicator
of the presence of high excitation optical emission lines, indicative of
the presence of an optical AGN. As in the GPS and CSS sources, the emission
lines are often broad with complex structure. There is evidence for
jet-cloud interactions in the form of blue-shifted lines in some of the
sources. Finally, a significant fraction of the sample show post-starburst
stellar continua.

We note that our galaxies have been selected to to have a clear mid-IR
excess rather than a radio excess, so it is perhaps not surprising that
classical optical signatures of radio jets interacting with the ISM are
largely missing. It remains to be understood if the radio emission in the
mid-IR excess AGN may be stellar in origin.  We also note that the
redshifts of the radio-excess IRAS AGN and the CSS/GPS sources studied in
the literature are in general higher than the galaxies in our two samples.

\subsection{Type II Quasars} Type II quasars are the obscured counterparts
of the classical quasar population predicted by AGN unification models.  In
the optical, Type II quasar candidates are traditionally selected as
objects with narrow permitted emission lines and high ionization line
ratios. Zakamska et al (2003) were the first to compile large  samples of
Type II quasar candidates in the Sloan Digital Sky Survey.  Their objects
were selected to lie in the redshift range $0.3 < z < 0.83$ in order to
disfavour selection of low luminosity objects. The  main disadvantage of
applying such a redshift cut in SDSS is that the Type II quasars are
not selected from a magnitude-limited survey of galaxies, so their
demographics and contribution to black hole growth cannot be studied in
detail, nor can direct comparisons be made to other classes of AGN.

Zakamska et al's  Type II  selection was based on a cut in the [OIII]/H$\beta$ line ratio
as well as the presence of very high-ionization lines such as [NeV]. The
[OIII] line luminosities of the sample range from $3 \times 10^7 L_{\odot}$
to close to $10^{10} L_{\odot}$ and the [OIII]/[OII] line ratios lie in the
range 1-10, i.e. very similar to the most luminous objects in our sample.
In follow-up work, Zakamska et al (2004) found that 143 of these objects
had counterparts in the FIRST radio catalogue. They speculate that this may
represent an overestimate of the true fraction, because the SDSS targetted
many FIRST radio sources for spectroscopy.  Hubble Space Telescope
imaging of the host galaxies of a subset of 9 Type II quasars with [OIII]
line luminosities greater than $3 \times 10^8 L_{\odot}$ reveal  that 6 out
of the 9 are elliptical galaxies well-fit by de Vaucouleurs light profiles
and the other 3 have a minor disk component (Zakamska et al 2006).  Most recently, Liu et al
(2013a,b) have obtained IFU data for 11 of the  most luminous, radio-quiet
objects in their sample and show the the [OIII] emission is very extended
with a mean diameter of 28 kpc and is spherical in morphology.  The
majority of nebulae show blue-shifted excesses in their line profiles across
most of their extents, signifying gas outflows. These authors estimate a
median outflow velocity of 760 km/s, similar to or above the escape
velocities from the host galaxies.

In Figure 15, we present a compilation of images of the mid-IR excess AGN
in our sample with the highest [OIII] luminosities ($> 10^8 L_{\odot}$). As
can be seen, the majority also have elliptical morphologies and centrally
concentrated light profiles.  The galaxy in the bottom row with a strange
red handle-liked protuberance is the famous ``teacup'' AGN, the nearest
known radio-quiet type II quasar with a redshift z=0.08056, first identified
by Reyes et al (2008). In recent work, Harrison
et al (2015) have studied the ionized gas kinematics in this object and
find evidence for an outflow with velocity 740 km/s. We thus believe it is
likely that there is a close correspondence between the Type II quasar
population and the brightest objects in our mid-IR excess sample.
We have also checked whether there is a significant population of 
optically-selected AGN with [OIII] luminosities  greater than ($> 10^8 L_{\odot}$)
that are not included in our mid IR-excess sample. We find only 25 out of 128 such
objects, indicating that the mid-IR and type II quasar
selection techniques yield essentially the same set of objects at the very
highest [OIII] luminosities.

\begin{figure*}
\includegraphics[width=135mm]{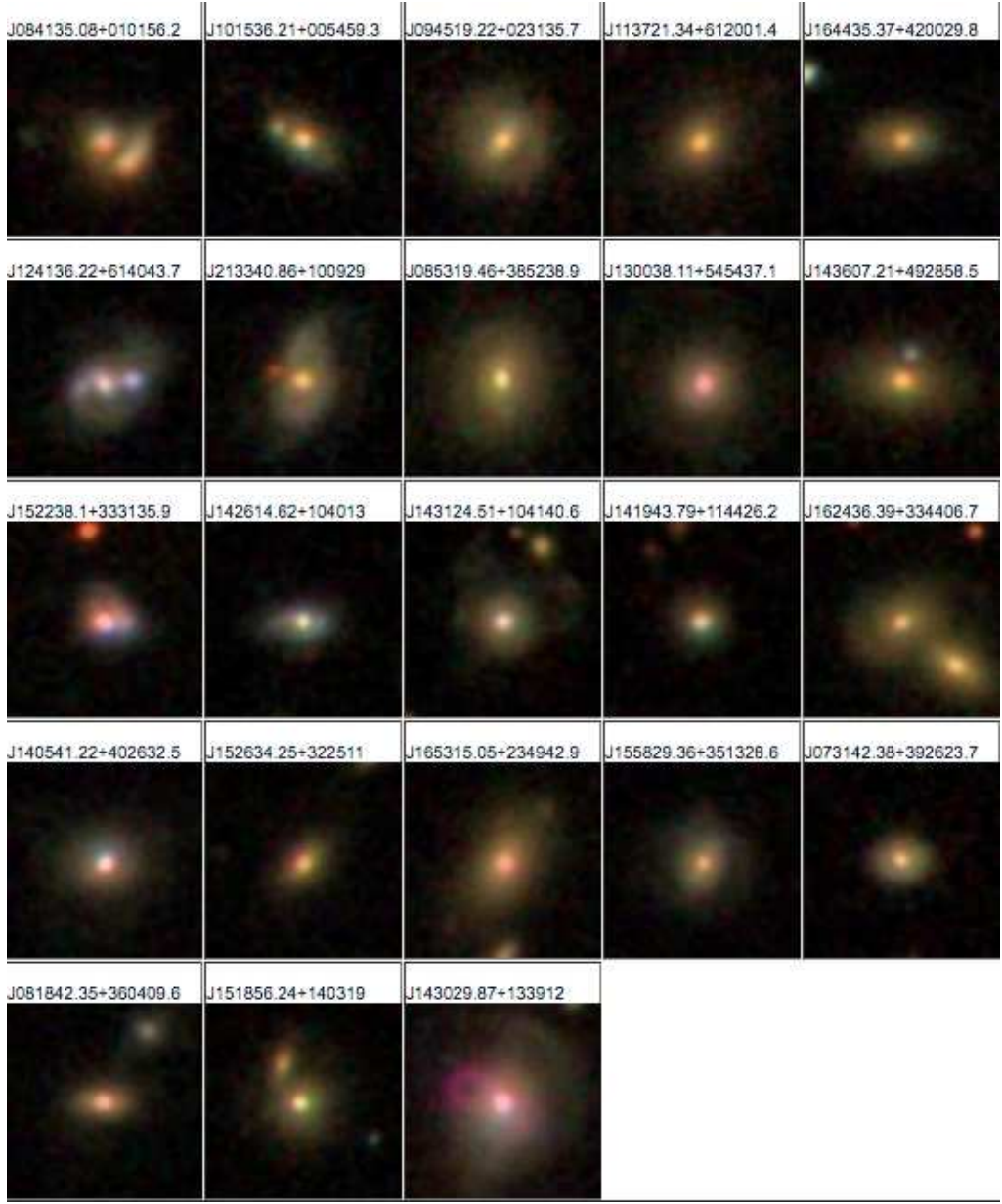}
\caption{A compilation of SDSS cut-out images of the mid-IR excess AGN
in our sample with the highest [OIII] luminosities ($> 10^8 L_{\odot}$). 
\label{models}}
\end{figure*}

What about the less luminous mid-IR excess AGN? In Figure 16, we
present a compilation of objects with D$_n$(4000) in the range 1.0-1.2,
indicative of current bursts of star formation. As shown in Figures 9 and
10, these galaxies have more moderate [OIII] luminosities in the range
$10^7-10^8 L_{\odot}$.  As can be seen, there are many more interacting
pairs and triples in this sample than in the high-luminosity sample.  If
the galaxies in Figures 15 and 16 constitute different phases of the same
merger-induced black hole fuelling events, the host galaxies shown in
Figure 16 could be said to be  an earlier stage of the merging process.

\begin{figure*}
\includegraphics[width=135mm]{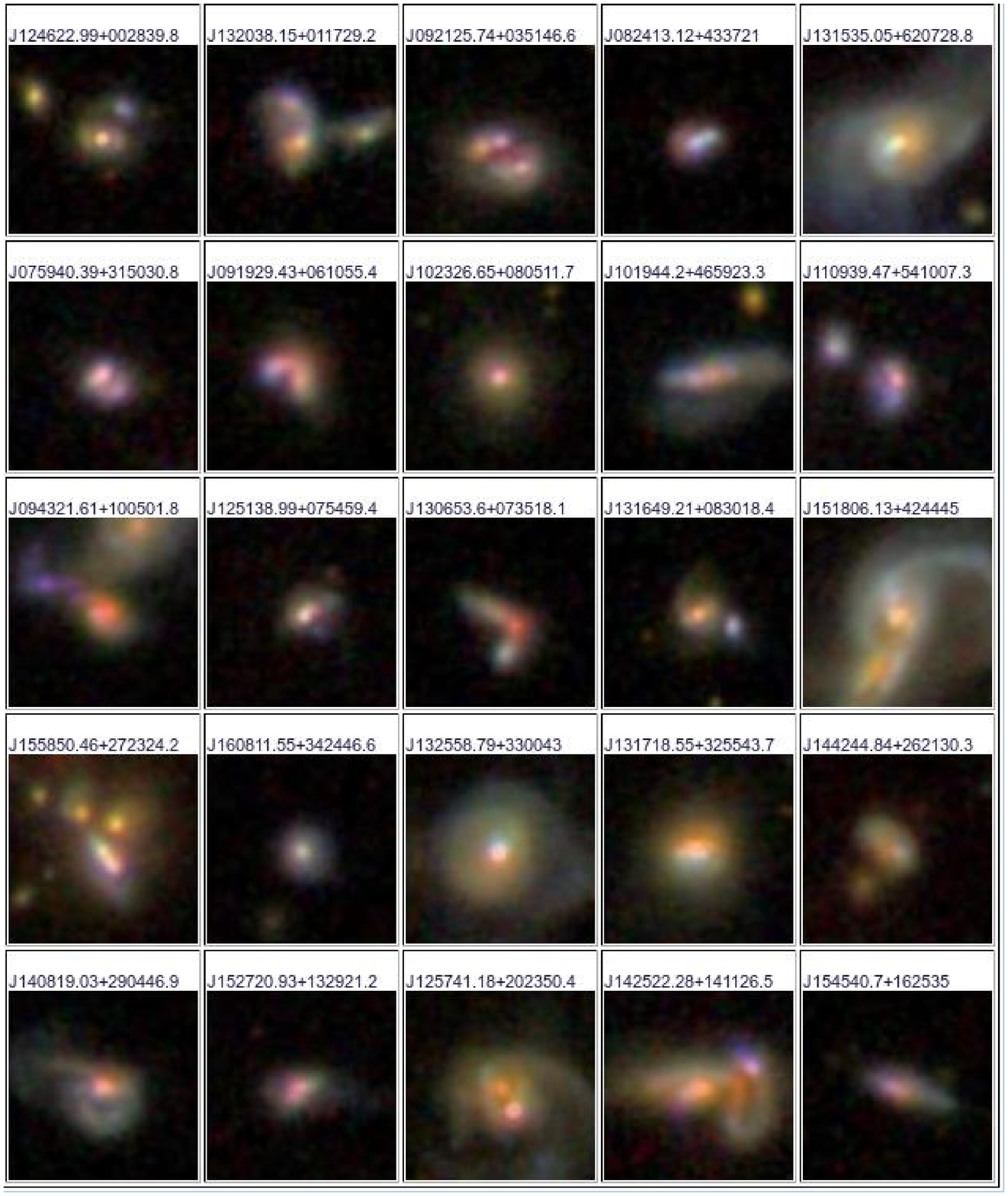}
\caption{A compilation of SDSS cut-out images of the mid-IR excess AGN
in our sample with D$_n$(4000) in the range 1.0-1.2, indicative
of current bursts of star formation. 
\label{models}}
\end{figure*}

\subsection {AGN selected only by a mid-IR colour criterion}
As discussed in Paper I, the selection of AGN using WISE photometry has generally
been carried out using colour cuts designed to avoid the main locus of star-forming
galaxies (e.g. Stern et al 2012). We showed in this paper that the W1-W2 colours
of a significant fraction of such objects remain very red out to large ($>$ 5 kpc)
radii, suggesting that a significant fraction of the mid-IR emission may arise
from an extended distribution of dust in the galaxy heated by collisions with
electrons from  surrounding hot halo gas, rather from a central, parsec-scale
torus. We include an additional radio-loud criterion
to increase the likelihood that the sample contains a high proportion of 
galactic nuclei with black holes that are currently accreting.  
In other words, our  goal is to maximize the purity of the 
AGN sample with the data at hand.

The danger with the procedure adopted in Paper I with respect to past
mid-IR selection procedures, is that such an AGN sample is not complete.
This may occur, for example, if the emission comes from a jet that is variable over timescales
that are short compared to the lifetime of the torus.   
Figure 17 compares some of the key properties of the full mid-IR excess sample
and the sample with the additional cut on radio luminosity. 
Mid-IR excess galaxies are  identified as outliers in SFR/$M_*$ versus D$_n$(4000)
space for galaxies with $\log$ SFR/$M_* > -11$ and in H$\delta_A$ versus D$_n$(4000) space
for galaxies with lower
specific star formation rates. We bin up the two planes in
intervals of 0.15 in D$_n$(4000), 0.25 dex in $\log$ SFR/$M_*$ and 0.1 in H$\delta_A$
and calculate the distribution of W1-W2 colours in
each bin. Outliers or mid-IR excess galaxies are defined to have colours
that lie above the upper 
95th percentile point of the distribution.
In Figure 17, we plot galaxy properties as a function of the quantity
$\Delta$(W1-W2), the difference between the measured W1-W2 colour of the galaxy and the
colour that delineates the upper 95th percentile cut.

In the top left panel, we  plot the fraction of galaxies as a function of
$\Delta$(W1-W2) for the mid-IR excess sample without the radio-loud cut (black
histogram) compared to the fiducial sample (black
triangles). As can be seen, the sample with the radio loud cut includes a
much more pronounced tail of objects with large $\Delta$(W1-W2), i.e. with
mid-IR colours that are very far from the stellar locus. This
supports our claim that the radio selection is increasing the purity of the
AGN sample. 

\begin{figure}
\includegraphics[width=90mm]{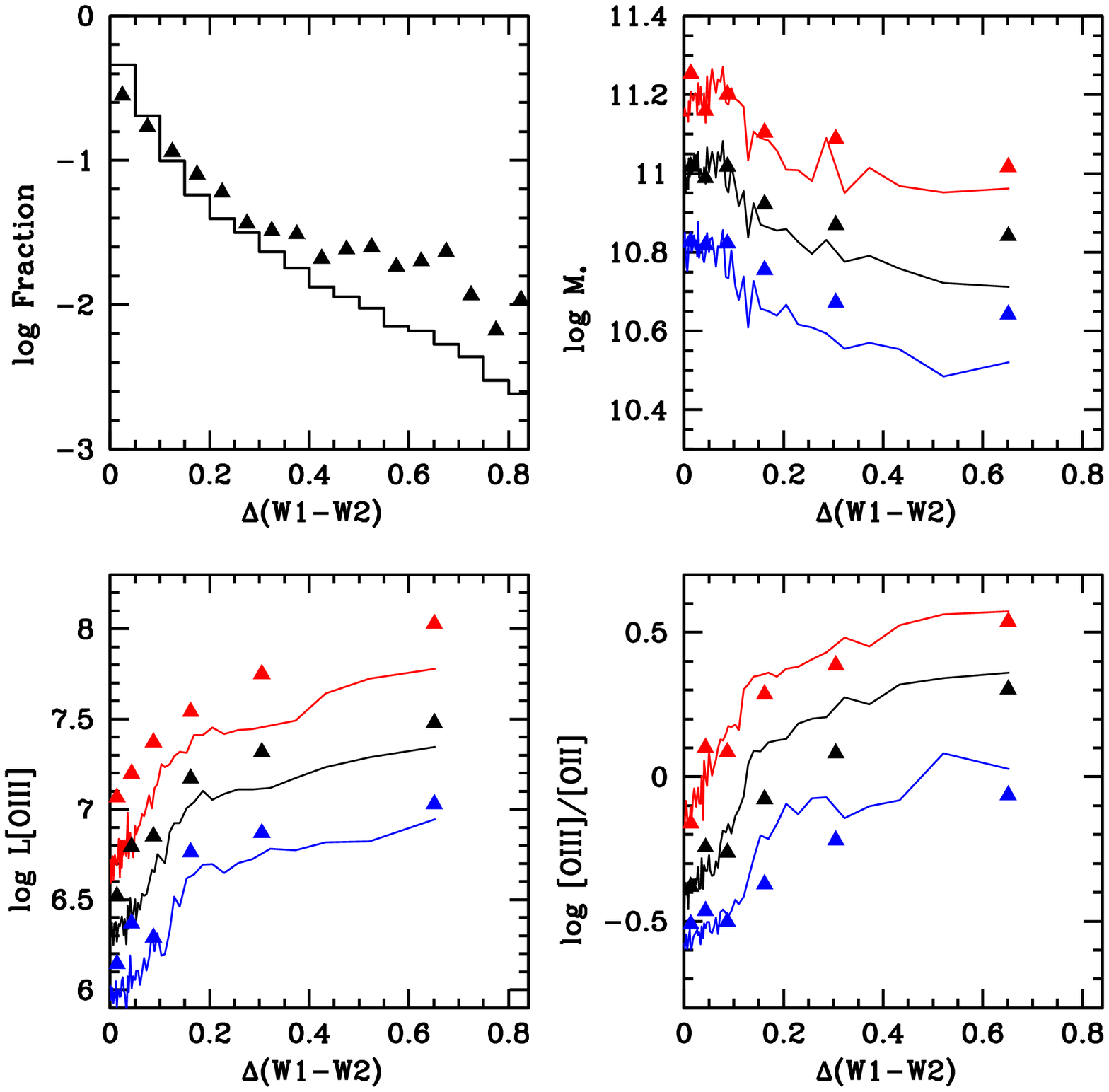}
\caption{ The top left panel shows the fraction of the sample
in bins of $\Delta (W1-W2)$, where $\Delta (W1-W2)$  is the difference between the W1-W2 colour and the
colour that delineates the 95th percentile cut.
The result for the mid-IR excess sample without any constraint from
the radio is shown as a black histograms, while black triangles show
the result for the fiducial sample.  
In the next three panels,  stellar mass, [OIII] line luminosity and
ionization parameter 
are plotted as a function of $\Delta (W1-W2)$. 
Each plotted point
corresponds to a bin containing a fixed number (200) galaxies, so
the noise due to Poisson sampling of the underlying distribution
remains constant in each diagram. Black line/triangles  show the median
of the distribution, red lines/triangles  the upper 75th percentile of the
distribution and blue lines/triangles  the lower 25th percentile. Once again,
lines denote the full mid-IR excess sample, while triangles denote the 
fiducial sample.
\label{models}}
\end{figure}

In the next three panels, trends in stellar mass of the host
galaxy, [OIII] line luminosity and ionization parameter [OIII]/[OII] are
shown as a function of $\Delta$(W1-W2) for the two samples. As can be seen,
[OIII] luminosity and 
ionization parameter  increase for systems with larger W1-W2 colours in a similar
manner for both classes of object.
This implies  that the main effect of the radio
selection is to boost the fraction of galaxies with the most extreme W1-W2
colours, which are also the most optically luminous systems with the
highest ionization parameters.  More detailed studies using spatially
resolved data will be necessary to figure out  the physical nature
of the central radio emission (jet or central starburst) and whether
or not the radio source is influencing the structure of the dust emission
in the central regions of the galaxy.

\section {Summary} We now summarize the main results of our analysis.  We
have  studied the narrow emission line properties and stellar populations
of a sample of 1385 radio-detected, mid-IR excess AGN in order to
understand the physical conditions in the interstellar medium of these
objects. We compare these systems with a control sample of 50,000 AGN
selected by their optical emission line ratios that do not have a
significant mid-IR excess. Our main conclusions 
are the following: \begin{itemize} \item The mid-IR excess AGN populate the
high ionization branches of the [OIII]/H$\beta$ versus
[OI]/H$\alpha$/[SII]/H$\alpha$ BPT diagrams, whereas the control sample AGN
cluster near the star-forming locus and have lower ionization parameters on
average.  \item The mid-IR excess AGN have [OIII] luminosities that are an
order of magnitude large on average than the control sample AGN.  \item The
mid-IR excess AGN have higher electron densities, but similar metal
abundances to the control sample.  \item The H$\delta_A$ versus D$_n$(4000)
diagrams show that a much larger fraction of the host galaxies of mid-IR
excess AGN have experienced recent bursts of star formation. These recent
starburst galaxies have lower stellar metallicities and higher Mg/Fe
ratios.  \item The number densities of mid-IR excess AGN are a 1000 times
smaller than those of control sample AGN at low [OIII] luminosities ($\sim
10^{6} L_{\odot}$, but at the very highest [OIII] luminosities probed by
our sample ($\sim 10^{9} L_{\odot}$), mid-IR excess AGN become more
populous by a factor of 10.  \item  Mid-IR excess AGN contribute about half
the total present-day black hole growth in galaxies with  stellar
masses larger than $10^{11} M_{\odot}$, whereas control sample AGN are 
currently the dominant contributor in lower
mass systems.  \end {itemize}

It is well known that the AGN population evolves strongly to higher
luminosities at higher redshifts, and it is likely that AGN similar to the
mid-IR excess population studied in this paper become much more populous.
We note that more than 95\% of all AGN in the parent sample with
[OIII] luminosities greater than $10^{8}$ L$_{\odot}$ are included in 
the mid-IR excess/radio sample studied in this paper, suggesting that
the two selection techniques converge at the highest luminosities. 

The future usefulness of our low redshift sample will lie in spatially
resolved spectroscopic follow-up studies of various kinds, as in the
Harrison et al study of the teacup AGN. These studies are required in order
to understand how accretion onto the central supermassive black hole is
occurring, the physical origin and location of the very high ionization gas
in these systems, and the impact of the energetic processes occurring near
the black hole on the interstellar medium of the host galaxy. The
establishment of a technique that selects a {\em population} of AGN seen at
different phases along a starburst cycle is also interesting for more
detailed follow-up programs. Although `AGN feedback' in the form of
extended outflowing gas is now established in a variety of  AGN
sub-populations such as the most luminous Type II quasars and radio
galaxies, understanding the global ubiquity, energetics  and duty cycle of
the feedback  will require more carefully controlled statistical
approaches. 

Finally, our sample is an interesting one for understanding
the relation between AGN activity, galaxy-galaxy interactions, 
mergers between black holes, and the origin of powerful AGN driven outflows of gas. 
It is rather interesting that although the number densities of IR excess
and control sample AGN are very different, the integrated [OIII] emissivity
in both classes of objects peaks at a stellar mass of $\sim 10^{10.5} M_{\odot}$.
We note that this was shown for the AGN population as a whole in Heckman \& Kauffmann
(2004, see their Figure 4). This value  ($10^{10.5} M_{\odot}$) 
corresponds closely to the transition mass where the galaxy population switches
over from a blue, star-forming, disk-dominated population to a red, passive, bulge-dominated
one. Energetic feedback from AGN has been hypothesized to cause this transition,
but considerable uncertainty remains as to how this occurs in practice. Some models
assume that a feedback mode associated with radio galaxies accreting hot gas 
at the centers of massive dark matter halos is responsible
for this transition (Croton et al 2006), while others invoke quasar-driven feedback
triggered by galaxy-galaxy mergers (Hopkins et al 2006). Identification of
a population of very luminous AGN clearly associated with galaxy-galaxy mergers 
is the first step to answering this question empirically.

\section*{Acknowledgments}
I thank Patricia Sanchez-Blazquez and Tim Heckman for helpful discussions and Mazda Adli
for his support.


\end{document}